\documentclass[letterpaper, 10 pt, conference]{ieeeconf}

\newcommand{\mbb}{\mathbb}
\newcommand{\mbf}{\mathbf}
\newcommand{\mc}{\mathcal}

\newcommand{\horizon}{T}
\newcommand{\state}{x}
\newcommand{\bstate}{\mbf{\state}}

\newcommand{\control}{u}
\newcommand{\bcontrol}{\mbf{\control}}

\newcommand{\dyn}{f}

\newcommand{\numplayers}{N}
\newcommand{\feedback}{\gamma}
\newcommand{\feedbackset}{\Gamma}
\newcommand{\runningcost}{g}
\newcommand{\cost}{J}
\newcommand{\arbitrarycost}{\mc{\cost}}

\newcommand{\xdim}{n}
\newcommand{\udim}{m}
\newcommand{\xconstraint}{\mc{X}}
\newcommand{\uconstraint}[1]{\mc{U}^{#1}}
\newcommand{\xset}{\mbb{R}^\xdim}
\newcommand{\uset}[1]{\mbb{R}^{\udim^{#1}}}

\newcommand{\regularization}{\epsilon}
\newcommand{\tofmax}{{t'}}
\newcommand{\reachlevel}{\alpha}

\newtheorem{remark}{Remark}
\newtheorem{assumption}{Assumption}

\newtheorem{definition}{Definition}

\newtheorem{theorem}{Theorem}

\newcommand{\figref}[1]{Fig.~\ref{#1}}
\newcommand{\secref}[1]{Sec.~\ref{#1}}
\usepackage{graphicx}
\usepackage{adjustbox}
\usepackage{wrapfig}
\usepackage[labelfont=bf,figurename=Fig.,font=small]{caption}
\usepackage{subcaption}
\usepackage{lipsum}
\usepackage{amssymb}
\usepackage{amsmath}
\usepackage{siunitx}
\usepackage{mathtools}
\usepackage{xcolor}
\usepackage{hyperref}
\usepackage[ruled,vlined,linesnumbered]{algorithm2e}
\usepackage[noend]{algpseudocode}
\usepackage{ifthen}
\usepackage{dsfont}
\usepackage[backend=biber,style=numeric-comp,sorting=none]{biblatex}
\usepackage{balance}

\bibliography{books, papers}

\IEEEoverridecommandlockouts
\title{\LARGE \bf Approximate Solutions to a Class of Reachability Games}

\author{
David Fridovich-Keil and Claire J. Tomlin
\thanks{
D. Fridovich-Keil is with the Department of Aeronautics \& Astronautics, Stanford University. C. Tomlin is with the the Department of Electrical Engineering \& Computer Sciences, UC Berkeley. Correspondence to \href{mailto:david.fridovichkeil@stanford.edu}{\tt \small{david.fridovichkeil@stanford.edu}}.}%
\thanks{This research is supported by an NSF CAREER award, the Air Force Office of Scientific Research (AFOSR), NSF's CPS FORCES and VeHICaL projects, the UC-Philippine-California Advanced Research Institute, the ONR MURI Embedded Humans, a DARPA Assured Autonomy grant, and the SRC CONIX Center.}
}

\begin{document}

\maketitle
\thispagestyle{empty}
\pagestyle{empty}

\begin{abstract}
In this paper, we present a method for finding approximate Nash equilibria in a broad class of reachability games. These games are often used to formulate both collision avoidance and goal satisfaction. Our method is computationally efficient, running in real-time for scenarios involving multiple players and more than ten state dimensions. The proposed approach forms a family of increasingly exact approximations to the original game. Our results characterize the quality of these approximations and show operation in a receding horizon, minimally-invasive control context. Additionally, as a special case, our method reduces to local gradient-based optimization in the single-player (optimal control) setting, for which a wide variety of efficient algorithms exist.
\end{abstract}

\section{Introduction}
\label{sec:intro}

\begin{figure}
    \centering
    \includegraphics[width=\columnwidth, trim=10cm 0cm 9cm 0cm, clip=true]{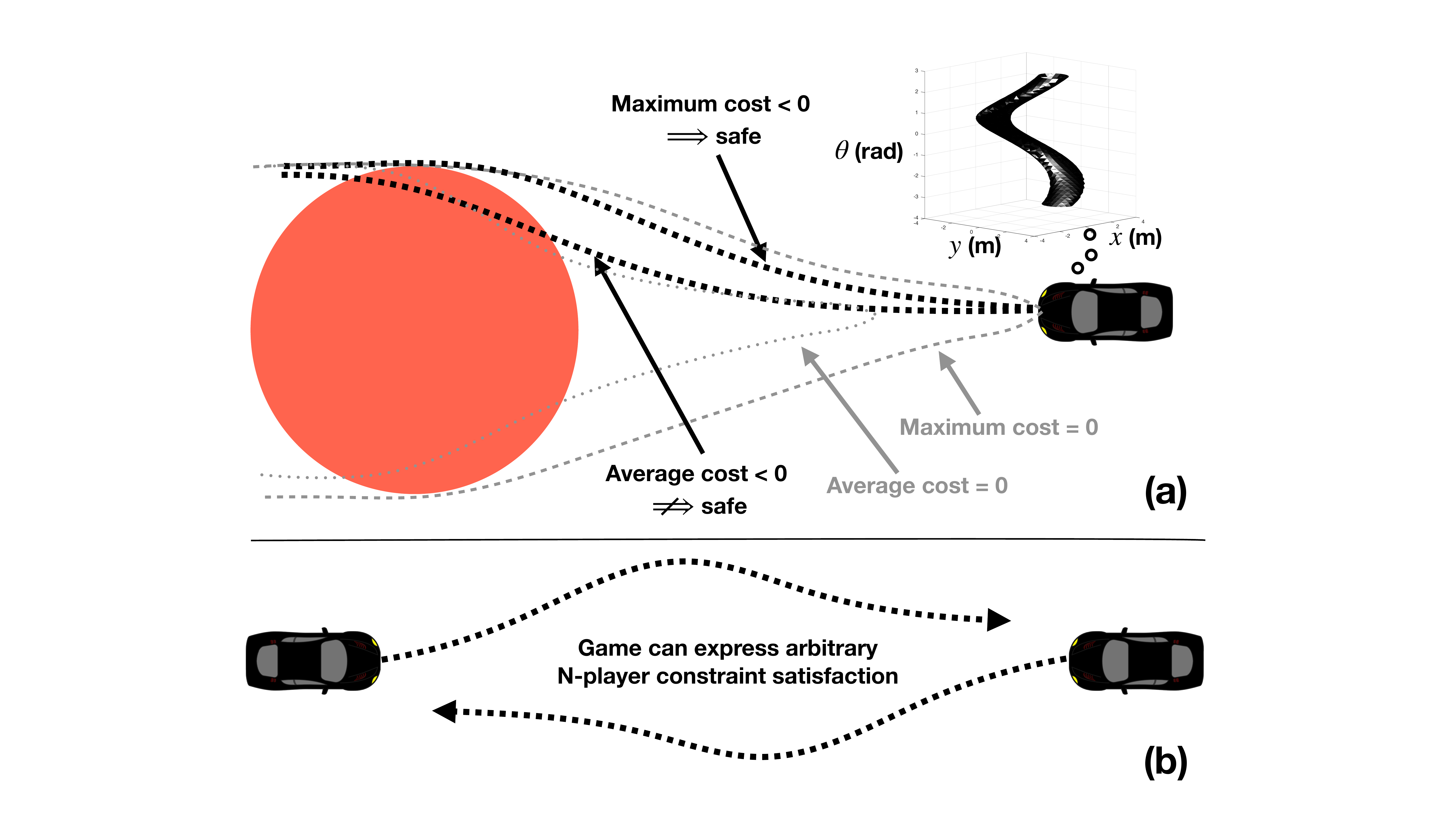}
    \caption{(a) The average cost of a trajectory does not imply safety (i.e., low maximum cost). Standard reachability methods can solve for an explicit \emph{safety boundary} where max cost is zero (plot shown above the vehicle). (b) Directly optimizing for the worst-case cost can guarantee collision-avoidance. This paper presents a scalable method for optimizing this worst-case in dynamic games.}
    \label{fig:front}
\end{figure}

Optimal control problems are often written with running, or time-additive, cost functions. That is, the objective of interest is typically a sum of time-varying functions over a fixed time horizon. Although this structure is reasonably general and easily amenable to both locally optimal and globally contractive optimal control methods, not all scenarios of interest can be expressed with a running cost. For example, in problems which encode properties like collision-avoidance (\figref{fig:front}), a time-additive objective can indicate safety even for an unsafe trajectory. Encoding these types of requirements with time-additive costs requires the introduction of (typically nonconvex) inequality \emph{constraints}, which complicate solution methods. On the other hand, by considering a maximum-over-time objective structure we can accurately assess the safety of trajectories without introducing explicit constraints. 
Similarly, a \emph{minimum}-over-time structure naturally expresses constraint satisfaction at \emph{any time}, rather than for all time. 
Aside from reducing problem complexity, these extremum-over-time formulations are also easily amenable to minimally-invasive control designs in which a nominal motion planner and tracking controller are used until a monitor detects potential constraint violation.


\figref{fig:front}a illustrates the importance of maximum-over-time objectives for encoding ``safety'' (understood as constraint satisfaction \emph{for all time}); one natural application is collision-avoidance. Encoding constraint satisfaction in this manner as an extremum-over-time is inherently a \emph{reachability} formulation. More generally,
reachability problems are concerned with whether a system always remains within (or ever enters) a ``target'' set. By contrast, other reachability problems are concerned with entering the target set at the final time and are expressed as terminal costs rather than extrema over time. 

In this paper, we consider the $N$-player general-sum dynamic game variant of these problems, in which each player may have an extremum-over-time cost. Our method finds approximate Nash equilibria of the game efficiently and in real-time. It relies upon making a family of approximations to the original game, which become increasingly precise and in the limit recover the original game.
Further, as a special case, the dynamic game formulation reduces to an optimal control problem in the single-player setting. Our method also applies here; however, we note that it is substantially similar to existing, well-developed optimization algorithms that can be readily applied in this setting.

The paper proceeds as follows. \secref{sec:background} provides a more formal description of the problem and a brief summary of the most related literature. \secref{sec:multi_player} presents our approach in the multi-player, general-sum game context, and \secref{sec:single_player} discusses its reduction to standard optimization techniques in the single-player optimal control setting. We conclude the paper with \secref{sec:conclusion} by noting several shortcomings of our method and interesting directions for future research. 

\section{Background}
\label{sec:background}

In this section, we shall present the core mathematical foundation of our approach and the corresponding related work. We shall treat the case of dynamic games separately from that of single-player optimal control, in which reachability problems are more traditionally studied. 
Further, we note that although some of the prior work we reference deals in continuous-time, our methods operate in discrete-time.

\subsection{Multi-Player Reachability Games}
\label{subsec:dynamic_games}

We address both the multi-player and the single-player settings in this paper. A multi-player dynamic game with $\numplayers$ players evolving over discrete-time $t \in \{1, \dots, \horizon\} \equiv [\horizon]$ is primarily defined by its dynamics, information pattern, and cost structure. 

The dynamics are specified as a difference equation, $\state_{t+1} = \dyn_t(\state_t, \control_t^{1:\numplayers})$, which describes the evolution of the state $\state_t \in \xconstraint_t \subseteq \xset$ with each player $i$'s control input $\control_t^i \in \uconstraint{i}_t \subseteq \uset{i}$. For clarity, in this work we we shall presume that these constraints are trivial, i.e. that $\state_t \in \xset, \control_t^i \in \uset{i}$.

The information pattern or \emph{strategy space} specifies what each player knows at each timestep. 
For our purposes, we shall presume a feedback structure in which each player knows the full state of the game $\state_t$ and chooses a corresponding input according to their strategy, i.e., $\control_t^i \equiv \feedback_t^i(\state_t), \feedback_t^i \in \feedbackset_t^i$ a measurable map from state to input for each player. 

Finally, the cost structure of the game may be defined arbitrarily for each player, i.e.,
\begin{align}
\cost^i(x_1, \feedback_{1:\horizon}^{1:\numplayers}) &\equiv \arbitrarycost^i(\bstate)\\ 
\label{eqn:game_cost_structure}
&:= \max_{t \in [\horizon]} \runningcost_t^i(\state_t)\,,
\end{align}
where we have defined $\bstate := \state_{1:\horizon+1}$, and $\runningcost_t^i(\state_t)$ encodes an arbitrary state cost at each time (such as distance-to-collision).
For clarity, we shall use maxima throughout the paper, although minima may also be used. 
We shall use the shorthand $\feedback^i \equiv \feedback_{1:\horizon}^i$ to represent each player's strategy over time. Additionally, note that player $i$'s cost function $\cost^i$ explicitly depends upon the initial condition $\state_1$ which determines the entire game trajectory $\bstate$. Further, note that we have presumed control-independence to restrict our attention to \emph{state} reachability problems.
For a much more detailed introduction to dynamic game theory, please refer to \cite{basar1999dynamic} and for their initial conception \cite{isaacs1954differential, isaacs1999differential}. 

Before proceeding, we note several important assumptions inherent to this formulation.
First, we have presumed a known, finite time interval over which the game takes place. Although this common assumption \cite{basar1999dynamic} appears to neglect a wide class of long-horizon and infinite-horizon problems, such issues are often practically addressed in optimal control settings via repeated solutions in a receding time horizon, e.g., \cite{borrelli2017predictive}.
Second, we have presumed that all players observe the full state of the game $\state_t$. Indeed, this is a strong assumption, especially in games with many players. Nevertheless, we assume full state feedback because it yields computationally-efficient solution methods such as \cite{fridovich2019efficient, fridovich2019iterative} and allows strategies to depend upon current information rather than only initial conditions as in other recent methods for solving smooth general-sum dynamic games \cite{cleac2019algames, di2019newton}. Partial observations are also possible in principle, but no efficient computational methods are known for this more general case. In the stochastic setting, however, recent results have shown promise \cite{schwarting2019stochastic}.

As in classical discrete games, smooth dynamic games admit a variety of solution concepts. We shall consider the well-known Nash equilibrium concept, in which we seek a set of strategies for all players, $\feedback^{1:\numplayers, *}$, where no player has a unilateral incentive to deviate from its strategy, i.e.
\begin{definition}{(Nash Equilibrium, see e.g. \cite[Chapter 6]{basar1999dynamic})}
\label{def:nash}
A set of strategies $\feedback^{1:\numplayers, *}$ is a Nash equilibrium if
\vspace{-0.2cm}
\begin{equation*}
    \cost^i\big(\state_1, \feedback^{1:\numplayers, *}\big) \le \cost^i\big(\state_1, \feedback^i, \feedback^{-i*}\big), \forall i \in [\numplayers], \forall \feedback_t^i \in \feedbackset_t^i\,,
    \vspace{-0.2cm}
\end{equation*}
where by $\feedback^{-i*}$ we denote the Nash strategies of all players other than $i$.
\end{definition}

\begin{assumption}
We shall presume the existence of a Nash equilibrium in the feedback strategies $\{\feedback^i\}$. Conditions under which this is guaranteed are detailed in \cite[Chapter 6]{basar1999dynamic}, but as a practical matter it is often possible to construct continuous dynamic games in which equilibria exist.
\end{assumption}

We note that, in a single-player optimal control setting where $\numplayers = 1$, a Nash equilibrium is precisely a globally optimal strategy. In this sense then, the method we develop for dynamic games also applies to optimal control problems. Since the single-player case is extremely well studied in its own right---and especially in the context of reachability---we present it separately. 

\subsection{Single-Player Setting}
\label{subsec:reach}

In the single-player case, our work can be understood as an approximate method for finding optimal trajectories for a class of Hamilton-Jacobi reachability problems \cite{mitchell2005time, fisac2015reach}. More precisely, we are concerned with choosing a discrete-time control signal which minimizes the extremum of a target function over time, and as before we restrict our attention to maxima although our contributions also apply for minima. As described above and illustrated in \figref{fig:front} this problem structure can be used to encode safety constraints.

These problems are characterized by \emph{dynamics} and \emph{cost structure}. The dynamics describe the evolution of the state variable $\state_t \in \xconstraint_t \subseteq \xset$ with control variable $\control_t \in \uconstraint{}_t \subseteq \uset{}$ as $\state_{t+1} = \dyn_t(\state_t, \control_t), \forall t \in [\horizon - 1]$. The cost structure and resulting optimal control problem are as follows:
\begin{align}
    \label{eqn:single_player_problem}
    \bstate^*, \bcontrol^* = \arg \min_{\bstate, \bcontrol} &\Big[\cost(\hat \state_1) = \max_{t \in [\horizon]}~\runningcost_t(\state_t)\Big]\\
    \textnormal{such that}~&\control_t \in \uconstraint{}_t, \forall t \in [\horizon]\,,\nonumber\\
    &\state_t \in \xconstraint_t, \forall t \in [\horizon]\,,\nonumber\\
    &\state_{t+1} = \dyn_t(\state_t, \control_t), \forall t \in [\horizon - 1]\nonumber\,,\\
    &\state_1 = \hat \state_1~\textnormal{(given initial state)}\nonumber.
\end{align}
with optimal trajectory given by $(\bstate^*, \bcontrol^*)$. Note that we use shorthand $\bstate := \state_{1:\horizon}, \bcontrol := \control_{1:\horizon}$ (in the multi-player context, we denote $\bcontrol^{1:N} \equiv \control_{1:T}^{1:N}$).

\begin{assumption}
We shall presume that feasible solutions always exist in \eqref{eqn:single_player_problem}. This assumption virtually always satisfied, e.g., if $\uconstraint{}_t$ is nonempty and $\xconstraint_t$ is sufficiently large.
\end{assumption}

This type of problem has been studied extensively in the literature. Broadly, approaches fall into three categories. (a) Conservative, often geometric, methods approximate the \emph{reachable set} which (depending upon the type of extremum) contains states from which either \emph{every} or \emph{at least one} trajectory ends in a given target set; examples include \cite{althoff2011zonotope, majumdar2014convex}. (b) Approximate dynamic programming methods (e.g., \cite{bertsekas2012approximate, bertsekas1996neuro}) attempt to find a parameterized ``value'' function which summarizes the cost-to-go from any state. (c)  Finally, grid-based Hamilton-Jacobi methods are one such method (summarized in \cite{bansal2017hamilton}), in which the value function is represented on a lattice. While resolution-complete, these methods suffer from Bellman's ``curse of dimensionality'' \cite{bellman1956dynamic} and do not generally scale to high-dimensional problems. 

Local methods for solving \eqref{eqn:single_player_problem} are also possible.
In the common setting of time-additive costs, gradient-based methods are both commonly-used and well-understood \cite{borrelli2017predictive}.
However, to the best of our knowledge gradient methods are not used for reachability problems with extrema-over-time objectives.
We show in \secref{sec:single_player} that, when applied in the single-player setting, our method for multi-player games reduces to just such a method.
Further, we show that a wide variety of other existing methods \cite{bertsekas1997nonlinear, nocedal2006numerical} may also be used in this case, thereby facilitating the further investigation of reachability methods in high-dimensional problems.

\section{Approximately Optimal Trajectories for Multi-Agent Reachability}
\label{sec:multi_player}

In this section, we consider an $\numplayers$-player dynamic game with maximum-over-time cost structure as in \eqref{eqn:game_cost_structure}, in which each player wishes to minimize this same form of objective, though now with different instantaneous costs $g^i$ for each player $i \in [\numplayers]$. Note that our approach also applies when some players instead have minimum-over-time or even sum-over-time objectives.

\subsection{Algorithmic Overview}
\label{subsec:implementation}

In practice, it is generally intractable to compute Nash equilibria \cite{papadimitriou2007complexity}. Instead, we settle for ``approximate local'' feedback Nash equilibria. These solutions are ``local'' in the sense that they only attempt to satisfy Def.~\ref{def:nash} within a small neighborhood in the strategy space \cite{ratliff2016characterization}, and ``approximate'' in the sense that they may still be a small distance $\delta$ from a local Nash equilibrium \cite{basar1999dynamic}. However, making these relaxations facilitates a much more computationally tractable algorithm for finding solutions---the iterative linear-quadratic (ILQ) method of \cite{fridovich2019efficient}.

This approach is similar to the ILQ regulator \cite{li2004iterative, todorov2005generalized, van2014iterated,  chen2017constrained} and differential dynamic programming \cite{mayne1966second, jacobson1970differential}, and generalizes the two-player zero-sum approach of \cite{mukai2000sequential}. Furthermore, it is worthwhile to note that the approach of \cite{fridovich2019efficient} finds approximate \emph{feedback} Nash solutions, which differ from the open-loop Nash solutions found in, e.g., \cite{di2019newton,cleac2019algames}.
Still, the quadratic approximation of cost developed in \secref{subsec:quadratic_approximation} is also compatible with open-loop methods \cite{di2019newton,cleac2019algames}.

For brevity, we omit a full description of the ILQ algorithm and direct the reader to \cite[Algorithm 1]{fridovich2019efficient}.
In short, however, the method iteratively refines strategies for all players using a closed form solution to a sequence of subproblems.
Each of these subproblems is formed by taking a linear approximation of game dynamics $\dyn_t$ and a quadratic approximation of each player's running cost $\runningcost^i_t$.
In \cite{fridovich2019efficient}, it is assumed that costs in these subproblems are time-additive; in this paper, we generalize to the case of extrema-over-time objectives found in reachability problems.

\subsection{Quadratic Approximation}
\label{subsec:quadratic_approximation}

The ILQ game approach of \cite{fridovich2019efficient} relies upon an efficient computation of the global feedback Nash equilibrium of a linear-quadratic game. Such an efficient solution is only known for games with sum-over-time structure  \cite{basar1999dynamic, starr1969nonzero}, not maximum-over-time problems as in \eqref{eqn:game_cost_structure}. To approximate a maximum-over-time cost structure in a sum-over-time form, we recognize that, at each iteration of the overall ILQ algorithm, a quadratic approximation to the maximum-over-time objective is entirely determined by the time at which maximum cost is achieved, i.e. $t = \tofmax$.
We consider variations $\state_\tofmax = \bar \state_\tofmax + h$ about state trajectory $\bar \bstate := \bar \state_{1:\horizon}$ and presume that $\runningcost_t^i(\state_t) < \runningcost_\tofmax^i(\state_\tofmax), \forall t \ne \tofmax$, which holds in nondegenerate cases. Thus equipped, we may form a Taylor expansion of the overall objective for player $i$ as
\begin{equation}
    \label{eqn:quadraticization}
    \arbitrarycost^i(\bstate) \approx \runningcost_\tofmax^i(\bar \state_\tofmax) + \frac{1}{2}\big(2 q_\tofmax^{iT} + h^T Q_\tofmax\big) h\,,
\end{equation}
where $q_\tofmax^i := \nabla_\state \runningcost^i_\tofmax(\bar\state_\tofmax)$ and $Q_\tofmax^i := \nabla^2_\state \runningcost^i_\tofmax(\bar\state_\tofmax)$.
Other terms in the expansion do not appear because, at $\bar \bstate$, overall cost $\arbitrarycost^i(\bar \bstate) = \max_{t \in [\horizon]} \runningcost^i_t(\bar \state_t) = \runningcost_\tofmax^i(\bar \state_\tofmax)$.
This implies that cost derivatives at other times (i.e., $q^i_t, Q^i_t, t \ne \tofmax$) are identically zero.

Thus, \eqref{eqn:quadraticization} constitutes a second-order Taylor expansion of player $i$'s maximum-over-time objective.
This quadratic approximation is trivially expressible as a sum-over-time in which all terms where $t \ne \tofmax$ are zero.
Hence, it is compatible with the ILQ game algorithm from \cite{fridovich2019efficient}.
As in the sum-over-time setting, this algorithm identifies \emph{approximate, local} feedback Nash equilibria. 

\subsection{Relaxation}
\label{subsec:relaxation}

Before applying the ILQ game method from \cite[Algorithm 1]{fridovich2019efficient}, we must ensure that solutions to the LQ game subproblems it solves at each iteration are well-defined. 
In state reachability games \eqref{eqn:game_cost_structure}, each player's objective does not depend upon its control input $\control_t^i$. However, analytical Nash solutions to LQ games require such dependence \cite[Chapter 6]{basar1999dynamic}; hence, we must also incorporate control dependence in each player's objective. 
That is, we must further approximate the instantaneous cost $\runningcost_t^i(\state_t)$ by adding a small but nonzero dependence upon control $\control_t^{1:\numplayers}$.
Concretely, we approximate each player's objective at level $\regularization > 0$ as follows:
\begin{align}
    \tilde \arbitrarycost^i_\regularization(\state_{1:\horizon}, \control_{1:\horizon}^{1:\numplayers}) &= \max_{t \in [\horizon]}~\tilde\runningcost_{t, \regularization}^i \nonumber\\
    \label{eqn:full_approx}
    &= \max_{t \in [\horizon]} \left\{\runningcost_t^i(\state_t) + \frac{1}{2}\regularization \|\control_t^i\|_2^2\right\}\,.
\end{align}

As we take $\regularization \to 0$, we recover the original problem under moderate assumptions. 

\begin{assumption}
\label{assume:existence}
Nash strategies for games with costs as in both \eqref{eqn:game_cost_structure} and \eqref{eqn:full_approx} exist and are feasible, for all positive $\regularization$.
\end{assumption}

\begin{assumption}
\label{assume:limit}
Limiting Nash strategies exist for the game with cost \eqref{eqn:full_approx} as $\regularization \to 0$ and are feasible.
\end{assumption}

\begin{theorem}
\label{theorem:control_limit}
If $\feedback^{1:\numplayers, *}$ are Nash strategies in the limit $\regularization \to 0$ for the game with relaxed objective~\eqref{eqn:full_approx}, then they are also Nash strategies in the original game with each players' cost as in \eqref{eqn:game_cost_structure}.

\begin{proof}
We rewrite player $i$'s problem in the original game with costs \eqref{eqn:game_cost_structure} as
\begin{align*}
    \feedback^{i*} = \arg \min_{\feedback^i} \max_{t \in [\horizon]} g_t^i(x_t) &= \arg \min_{\feedback^i} \lim_{\regularization \to 0}\max_t~\tilde \runningcost_{t, \regularization}^i \\
    &= \lim_{\regularization \to 0} \arg \min_{\feedback^i}\max_t~\tilde \runningcost_{t, \regularization}^i = \feedback^{i*}\,,
\end{align*}
where we can interchange the limit and minimum safely by Assumptions~\ref{assume:existence} and \ref{assume:limit}.

Thus, if these assumptions hold, the limiting Nash strategies in the relaxed game \eqref{eqn:full_approx} are Nash in the original game with player costs from \eqref{eqn:game_cost_structure}.
\end{proof}
\end{theorem}

\begin{remark}
We note that care must be taken in the context of zero-sum games.
Here, each LQ subproblem of \cite[Algorithm 1]{fridovich2019efficient} is especially susceptible to poor numerical conditioning due to the inherent adversarial competition of these games.
Although these issues may be addressed with further regularization, we do not treat them here.
Control constraints $\control^i_t \in \uconstraint{i}_t$ are often necessary in such settings; such constraints are neglected in this work but may be addressed using penalty and interior point methods.
\end{remark}

\subsection{Example: Three-Player Avoidance Game}
\label{subsec:example_game}

Here and in \secref{subsec:minimally_invasive} we demonstrate our method in operation.
As our method is the first to treat $\numplayers$-player general reachability games, a direct comparison to prior work is not possible.

Consider a three-player dynamic game where each player has bicycle dynamics, i.e.
\begin{equation}
    \label{eqn:bicycle}
    \dot \state^i = \frac{d}{dt}\begin{bmatrix}
        p_x^i\\
        p_y^i\\
        \theta^i\\
        \phi^i\\
        v^i
    \end{bmatrix} = \begin{bmatrix}
        v^i \cos(\theta^i)\\
        v^i \sin(\theta^i)\\
        v^i \tan(\phi^i) / L^i\\
        \omega^i\\
        \alpha^i
    \end{bmatrix}\,.
\end{equation}
Here, player $i$'s input $u^i = (\omega^i, \alpha^i)$ controls the front wheel angular rate ($\dot \phi^i$) and linear acceleration ($\dot v^i$), respectively. Each bicycle has inter-axle distance $L^i > 0$, which enforces a minimum positive turning radius.

We construct a dynamic game in which each player has Euler-discretized dynamics according to \eqref{eqn:bicycle} with a sampling time of $\SI{0.1}{\second}$ and time horizon $\SI{2}{\second}$ (used throughout), and the following cost structure (written for player 1 for brevity), which penalizes player 1 for the smallest relative distance between it and players 2 and 3:
\begin{align}
    \label{eqn:game_cost}
    \runningcost_t^1 = \bar{d}-\min\{&d_t^{12},~d_t^{13}\},\\
    \textnormal{where}~d_t^{12} &:= \|(p_{x, t}^1, p_{y, t}^1) - (p_{x, t}^2, p_{y, t}^2)\|_2\,,\nonumber\\
    d_t^{13} &:= \|(p_{x, t}^1, p_{y, t}^1) - (p_{x, t}^3, p_{y, t}^3)\|_2\,,\nonumber
\end{align}
and define $\tilde \runningcost_{t, \regularization}^i$ as in \eqref{eqn:full_approx}. Here, $\bar{d}$ is a desired separation between the vehicles.

\begin{figure}
    \centering
    \includegraphics[width=0.99\linewidth, trim=10cm 0cm 10cm 0cm, clip=true]{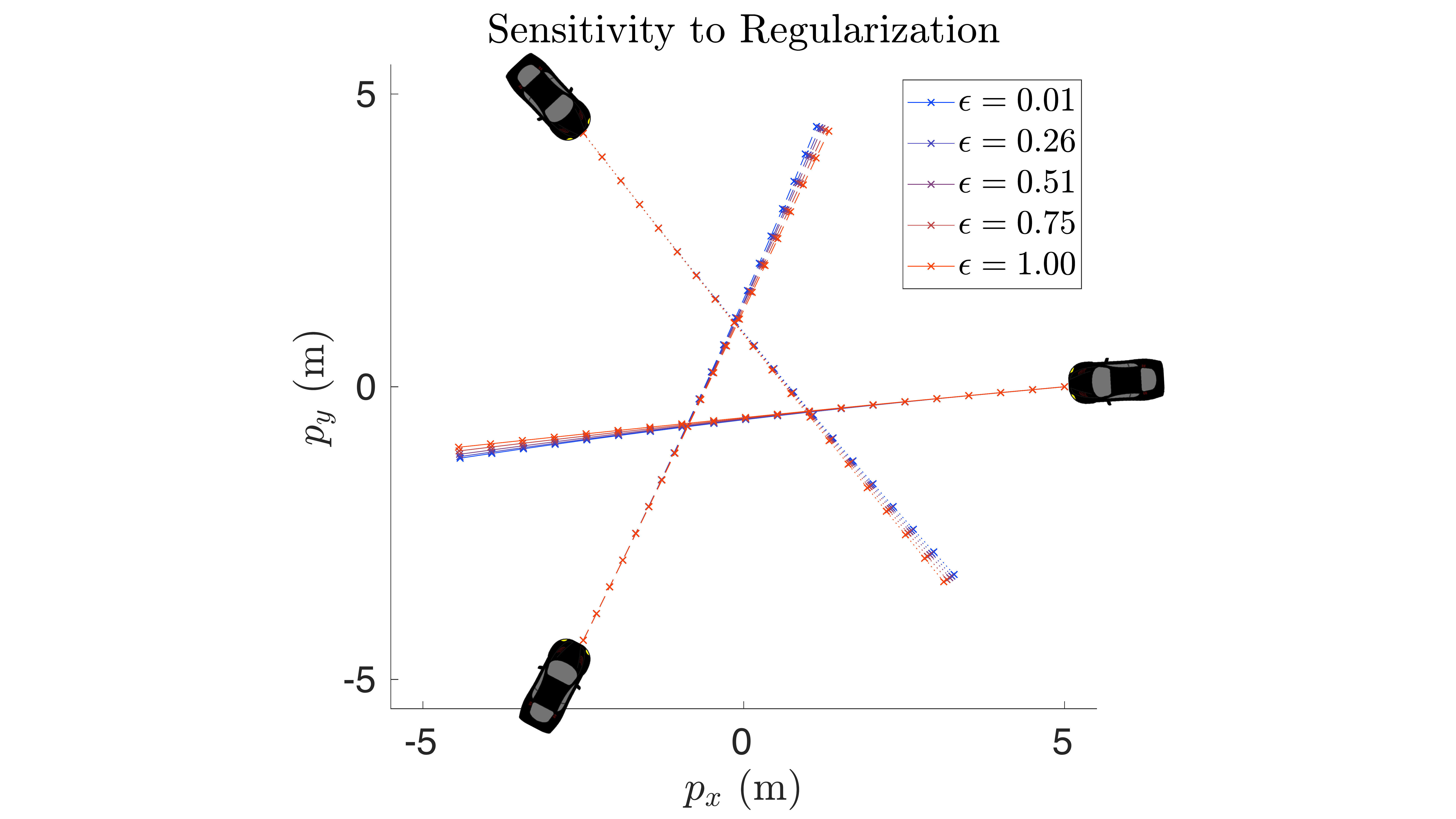}
    \caption{Optimal trajectories in the three-player avoidance game for varying $\regularization$ (black cars mark the initial position for each player). As $\regularization \to 0$ the more blue trajectories display a greater tendency toward avoidance.}
    \label{fig:game_avoid}
\end{figure}

We plot the approximately optimal strategies for the relaxed game constructed according to \eqref{eqn:full_approx} in \figref{fig:game_avoid}, showing the affect of decreasing $\regularization \to 0$. As we approach this limit, the players take more extreme avoidance maneuvers. Additionally, for each $\regularization$ our method finds a solution reliably in real-time under $\SI{1}{\second}$.

\begin{figure*}
    \centering
    \includegraphics[width=\linewidth, trim=0cm 0cm 0cm 0cm, clip=true]{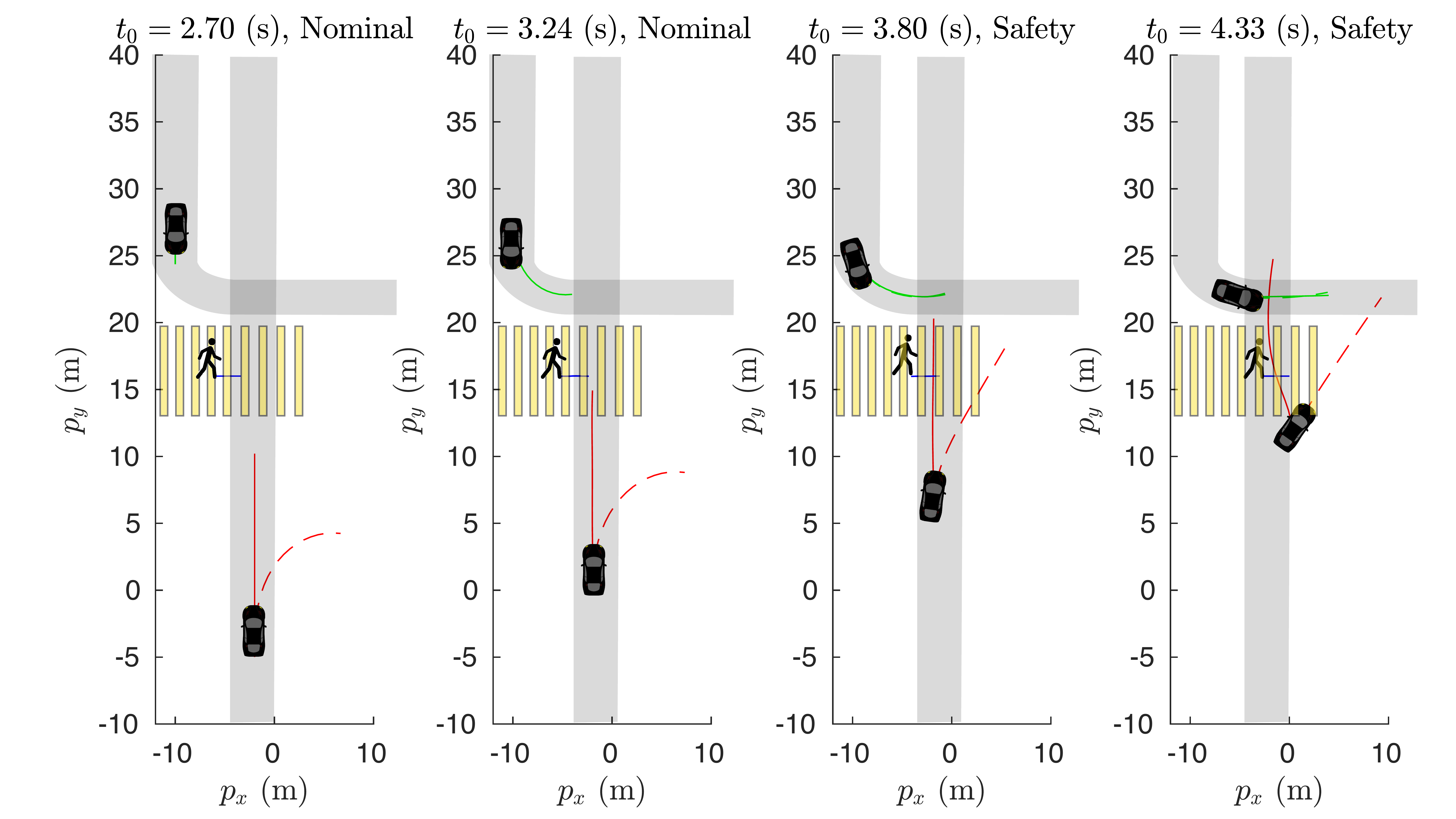}
    \caption{Optimal trajectories in the intersection game, executed with a receding time horizon. Solid lines indicate nominal trajectories. Dashed lines are safe trajectories for the ego vehicle, which successfully avoids collision by following a minimally-invasive control strategy.}
    \label{fig:minimally_invasive}
\end{figure*}

\subsection{Example: Receding Horizon, Minimally-Invasive Control}
\label{subsec:minimally_invasive}

Perhaps the most practical usage of reachability-based controllers is in receding horizon and minimally-invasive settings, where a single ``ego'' agent overrides its nominal controller whenever safety is nearly violated. In receding horizon problems, players' strategies at time $t$ only match those for the most recently solved game, not earlier ones; hence, collision-avoidance is not generally guaranteed. For a recent analysis of the induced information structure, please refer to \cite{petrosian2019hamilton}. \figref{fig:minimally_invasive} demonstrates our method's operation in a receding horizon, minimally-invasive setting for a three-player intersection game resembling that in \cite{fridovich2019efficient}.

Here, the ego vehicle (at bottom, red trajectory) and another car (both system dynamics are as above) navigate an intersection while a pedestrian crosses a crosswalk (its dynamics are those of a standard planar unicycle model). We set up costs for a nominal game as in \cite{fridovich2019efficient}, and in order to emphasize the role of a safety controller, we reduce the cost weight for maintaining sufficient proximity between agents. We also construct an identical reachability game to encode collision-avoidance, in which the ego vehicle's objective is of the form \eqref{eqn:game_cost}, and its equilibrium trajectory is shown in dotted red. Note that it avoids proximity to a much greater degree (i.e., behaves much more conservatively) than the nominal strategy; this is typical of safety strategies. As shown in \figref{fig:minimally_invasive}, over time the ego vehicle switches to the safety strategy as its planned (nominal) trajectory nears other agents. Time $t_0$ refers to the time of each planning invocation of horizon $\horizon = \SI{2}{\second}$. As above, our method operates in real-time, and since each receding-horizon invocation can be warm-started with the previous solution the amortized speed per invocation is typically on the order of $\SI{0.1}{\second}$ or less.

Computation of a Nash solution
is centralized rather than distributed among all agents.
As in robust planning methods, e.g. \cite{hardy2013contingency}, we envision such computation constituting the motion planning process for a single ego vehicle. Here, the reachability game formulation allows the ego vehicle to reason about how other vehicles may react to its own control decisions and thereby allow it to find interactive solutions in multi-agent settings.
\section{Approximate Optimal Trajectories for Single-Agent Reachability}
\label{sec:single_player}

The single-player reachability problem \eqref{eqn:single_player_problem} is already in the form of a nonconvex, nonlinear program.
In this case, our approach from \secref{sec:multi_player} may be viewed as a direct adaptation of well-known ILQ regulator of \cite{li2004iterative, todorov2005generalized, van2014iterated,  chen2017constrained} to the setting of state reachability.
However, a wider variety of methods for approximately solving general nonlinear programming problems can be found, e.g., in \cite{bertsekas1997nonlinear, nocedal2006numerical}.
Here, solution methods are approximate in the sense that they find local optima. 
Due to this approximation, solutions are \emph{conservative} as discussed below. 

\begin{theorem}
\label{thm:conservative}
Define optimal cost or \emph{value} $V(\hat \state_1)$ as the globally minimum cost $\cost^*(\hat \state_1)$ attained by \eqref{eqn:single_player_problem} at initial state $\hat \state_1$, and $\tilde V(\hat x_1)$ a local minimum identified by a particular nonlinear programming algorithm. For any scalar threshold $\reachlevel \in \mbb{R}$, the following inequalities hold:
\begin{align}
    \cost^*(\hat \state_1) \equiv V(\hat \state_1) &\le \tilde V(\hat \state_1)\,,~\textnormal{and}\\
    \{\hat \state_1 : V(\hat \state_1) \le \reachlevel\} &\subseteq \{\hat\state_1 : \tilde V(\hat\state_1) \le \reachlevel\},~\forall \hat \state_1 \in \xconstraint_1\,.
\end{align}
Sublevel sets of this form are called \emph{reachable sets} and these inequalities imply that the solution is \emph{conservative}, i.e., any state it concludes is safe (with sufficiently small cost) would also be safe for the globally optimal controller.

\begin{proof}
The first inequality is a consequence of the local optimality of solutions to nonconvex programming, and the second inequality follows from the first.
\end{proof}
\end{theorem}

Further, in the special case in which \eqref{eqn:single_player_problem} is convex, we have the following result.

\begin{theorem}
\label{thm:exact_if_convex}

Define $V$ and $\tilde V$ as before, and suppose that problem \eqref{eqn:single_player_problem} is convex in the decision variables, i.e., that sets $\xconstraint_t, \uconstraint{}_t$ are convex for all $t$, dynamics $f$ are affine in $\state_t, \control_t$ for all $t$, and instantaneous cost $\runningcost_t$ is a convex function of its argument (and that the cost has a maximum-over-time structure). Then, the local minimum cost $\tilde V$ is equivalent to the globally minimal $V$.

\begin{proof}
Follows from the global optimality of solutions to convex programs.
\end{proof}
\end{theorem}

Our main algorithmic contribution treats the multi-player noncooperative game setting.
While our method applies in this single-player setting, to demonstrate the general utility of posing single-player reachability problems as nonlinear programs we utilize existing, off-the-shelf methods instead.
That is, by resorting to local methods for nonlinear programming, solution methods may be significantly more computationally efficient than global techniques such as \cite{mitchell2005time, majumdar2014convex}. More to the point, to our knowledge, these methods are not widely used in the reachability literature and have the potential to dramatically improve practical performance in realistic, high-dimensional cases.

\subsection{Example: High-Dimensional Quadrotor}
\label{subsec:quad}

To demonstrate the computational advantages of local solutions for single-player reachability problems, we construct a high-dimensional quadrotor example with a 14-dimensional state space. The continuous-time dynamics are given below and may be found in \cite{al2009quadrotor}:
\begin{align}
    \state &= (p_x, p_y, p_z, \psi, \theta, \phi, v_x, v_y, v_z, \zeta, \xi, p, q, r)\nonumber\\
    \label{eqn:quad}
    \frac{d}{dt}&\begin{bmatrix}
        p_x\\
        p_y\\
        p_z\\
        \psi\\
        \theta\\
        \phi\\
        v_x
    \end{bmatrix} = \begin{bmatrix}
        v_x\\
        v_y\\
        v_z\\
        q\\
        r\\
        p\\
        g_x \zeta
    \end{bmatrix},~ 
    \frac{d}{dt}\begin{bmatrix}
        v_y\\
        v_z\\
        \zeta\\
        \xi\\
        p\\
        q\\
        r
    \end{bmatrix} = \begin{bmatrix}
        g_y \zeta\\
        g_z \zeta - g\\
        \xi\\
        \tau\\
        \alpha_x / I_x\\
        \alpha_y / I_y\\
        \alpha_z / I_z
    \end{bmatrix},\\
    \textnormal{with}~g_x &= (\sin(\phi)\sin(\psi) + \cos(\phi)\cos(\psi)\sin(\theta)) / m,\nonumber\\ 
    g_y &= (\cos(\phi)\sin(\psi)\sin(\theta) - \cos(\psi)\sin(\phi)) / m,\nonumber\\
    g_z &= \cos(\phi)\cos(\theta) / m\nonumber\,, 
\end{align}
where $g=\SI{9.81}{\meter \per \square \second}$ is the acceleration due to gravity. Here, $(m, I_x, I_y, I_z)$ are mass and inertia parameters (in our example, set to unity), and the states include positions, angles, their derivatives, and a double integrator on thrust. The controls are the second-derivative of thrust ($\tau$), and angular accelerations $(\alpha_x, \alpha_y, \alpha_z)$.
As before, we use an Euler discretization.

\begin{figure}
    \centering
    \includegraphics[width=0.99\columnwidth, trim=2cm 6.5cm 2cm 6.25cm, clip=true]{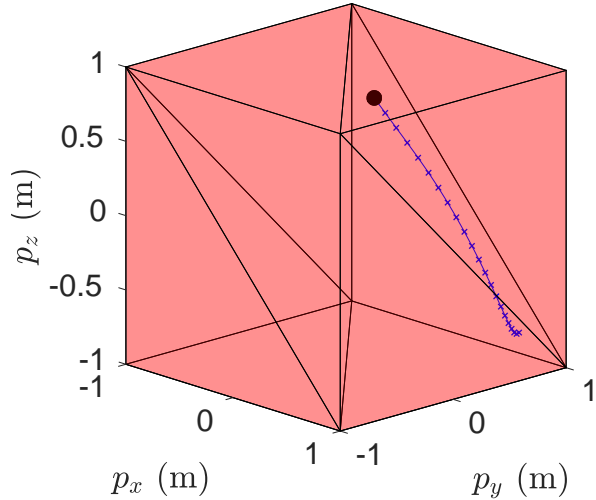}
    \caption{Locally optimal trajectory for high-dimensional quadrotor. The boundary of the unsafe region where $g > 0$ is shown in red. The solid black dot denotes the quadrotor's initial position.}
    \label{fig:quad_avoid}
\end{figure}

The instantaneous cost records the signed distance from the boundary of a cube, i.e. $\runningcost(x) = \|(p_x, p_y, p_z)\|_\infty - B$. Marking the boundary of the unsafe region in which $\runningcost > 0$ in red, \figref{fig:quad_avoid} displays the approximately optimal trajectory. 
To emphasize the generality of the nonlinear programming formulation and the quality of existing solution methods in this single-player context, we use well-known existing tools such as IPOPT \cite{biegler2006implementation} and SNOPT \cite{GilMS05, snopt77}, accessed via YALMIP \cite{Lofberg2004} in MATLAB\textsuperscript{\textregistered}.

\section{Discussion}
\label{sec:conclusion}

This paper presents a novel approach to solving extremum-over-time $\numplayers$-player reachability games. Our method is computationally tractable and yields real-time approximate solutions to high-dimensional problems. In the game setting, we introduce an arbitrarily accurate relaxation of the original objective structure and provide a guarantee of limiting convergence, and in the single-player optimal control setting, we provide a conservativeness guarantee. 
We have demonstrated our approach in several examples, including a three-player dynamic game, a receding horizon traffic example, and a high-dimensional quadrotor example. In the remainder of this section, we shall outline promising extensions and directions for future work.

\subsection{Numerical Stability}
\label{subsec:numerical_stability}

The choice of $\regularization$ may have significant impact upon the numerical stability of the ILQ algorithm, since the underlying LQ feedback Nash solution relies upon solving linear systems of equations which become singular when $\regularization \approx 0$. Further work is needed to improve numerical stability in these cases. 

\subsection{Constraints}
\label{subsec:constraints}

A related, promising direction is the incorporation of constraints on both states and inputs. Constrained games require a distinct notion of equilibrium, such as that of generalized Nash equilibria. Our current work investigates efficient methods for solving these types of feedback games, although it is worth noting that solutions already exist for the simpler \emph{open-loop} information pattern \cite{cleac2019algames}.

\subsection{Annealing}
\label{subsec:annealing}

Although we have not investigated it in this work, we believe that annealing $\regularization \to 0$ may improve asymptotic approximation quality. Such annealing may encourage the iterative ILQ method of \secref{subsec:implementation} to find globally optimal solutions of the original reachability problem. We note that, because each subproblem with fixed $\regularization$ may be solved so rapidly, the computational burden of annealing is minimal.




\balance
\printbibliography

\end{document}